\documentclass[prd,aps,nofootinbib,preprint,tightenlines]{revtex4}

\newcommand{\checked}[1]{}

\usepackage{graphicx}

\newcommand{\beq}{\begin{equation}}
\newcommand{\eeq}{\end{equation}}
\newcommand{\bea}{\begin{eqnarray}}
\newcommand{\eea}{\end{eqnarray}}
\def\simge{\mathrel{
  \rlap{\raise 0.511ex \hbox{$>$}}{\lower 0.511ex \hbox{$\sim$}}}}
\def\simle{\mathrel{
  \rlap{\raise 0.511ex \hbox{$<$}}{\lower 0.511ex \hbox{$\sim$}}}}

\begin{document}

\title{Quarkonium states in an anisotropic QCD plasma}


\author{Adrian Dumitru$^{a,b,c}$, Yun Guo$^{d,e}$, \'Agnes M\'ocsy$^f$ and 
Michael Strickland$^{g}$}
\affiliation{$^a$Department of Natural Sciences, Baruch College, CUNY,
17~Lexington Ave, New York, NY 10010, USA\\
$^b$RIKEN-BNL Research Center, Brookhaven National Lab, Upton, NY
11973, USA\\
$^c$Graduate School and University Center, City University of New
York, 33 W 42nd St., New York, NY 10036, USA\\
$^d$Helmholtz Research School,
Johann Wolfgang Goethe Universit\"at,
Max-von-Laue-Str.\ 1, D-60438 Frankfurt am Main, Germany\\
$^e$Institute of Particle Physics, Huazhong Normal University,
Wuhan 430079, China\\
$^f$Department of Mathematics and Science, Pratt Institute,
200 Willoughby Avenue,
Brooklyn, NY 11205, USA\\
$^g$Department of Physics, Gettysburg College, Gettysburg, PA 17325, USA
\vspace*{2cm}
}

\begin{abstract}
  We consider quarkonium in a hot QCD plasma which, due to expansion
  and non-zero viscosity, exhibits a local anisotropy in momentum
  space. At short distances the heavy-quark potential is known at tree
  level from the hard-thermal loop resummed gluon propagator in
  anisotropic perturbative QCD. The potential at long distances is
  modeled as a QCD string which is screened at the same scale as the
  Coulomb field. At asymptotic separation the potential energy is
  non-zero and inversely proportional to the temperature. We obtain
  numerical solutions of the three-dimensional Schr\"odinger equation
  for this potential. We find that quarkonium binding is stronger at
  non-vanishing viscosity and expansion rate, and that the anisotropy
  leads to polarization of the $P$-wave states.
\end{abstract}
\maketitle
\newpage

\section{Introduction}

In the era of the Relativistic Heavy Ion Collider at Brookhaven and
the Large Hadron Collider at CERN the theoretical understanding of
in-medium modifications of QCD bound states is expected to
progress significantly. In this paper we focus on the properties of
bound states of heavy quarks (quarkonia) in an anisotropic plasma.

In Quantum Chromodynamics (QCD) with small t' Hooft coupling at short
distances non-relativistic quarkonium states exist. Their binding
energies are much smaller than the quark mass $m_Q\gg\Lambda_{\rm
  QCD}$ ($Q=c,b$), and their size is much larger than $1/m_Q$. At zero
temperature, since the velocity of the quarks in the bound state is
small, $v\ll 1$, quarkonium can be understood in terms of
non-relativistic potential models \cite{Lucha:1991vn} using the
Cornell potential \cite{Eichten:1979ms}. The potential model can
actually be derived directly from QCD as an effective field theory
(potential non-relativistic QCD - pNRQCD) by integrating out modes
above the scales $m_Q$ and then $m_Q v$, respectively
\cite{Brambilla:2004jw}.

At high temperatures, the deconfined phase of QCD exhibits screening
of static color-electric fields~\cite{GPY}. It is expected that this
screening leads to the dissociation of quarkonium states, which can
serve as a signal for the formation of a deconfined quark-gluon plasma
in heavy ion collisions~\cite{MatsuiSatz}. Inspired by the success at
zero temperature, potential model descriptions have also been applied to
understand quarkonium properties at finite temperature. The pioneering
paper of Matsui and Satz \cite{MatsuiSatz} was followed by the work of
Karsch, Mehr and Satz (KMS) \cite{KMS}, which presented the first
quantitative calculation. In recent works more involved calculations
of quarkonium spectral functions and meson current correlators
obtained from potential models have been performed
\cite{mocsy,wong,alberico,rapp,Mocsy:2007jz,Mocsy:2007yj}. The results
have been compared to first-principle QCD calculations
performed numerically on lattices \cite{datta,jakovac,aarts,umeda}.  A
summary and review of the current understanding of these potential
models is presented in \cite{Mocsy:2008eg}, and different aspects of
quarkonium in collider experiments can be found in \cite{LHCpred}.
More recently the imaginary part of the potential due to Landau
damping has been calculated \cite{laine,blaizot}. Also, the derivation
of potential models from QCD via effective field theory methods has
been extended to finite $T$~\cite{Brambilla:2008cx}. All of these
works, however, have been performed with the assumption of an
isotropic thermal medium.

Here, we attempt a first assessment of the properties of quarkonium
states in a QCD plasma which exhibits an anisotropy in momentum
space. Such anisotropy may arise due to a locally anisotropic
hydrodynamic expansion of a plasma with non-vanishing shear viscosity. It
leads to an angular dependence of the $Q\overline{Q}$
potential~\cite{Dumitru:2007hy}. We note that the non-equilibrium
effect described here arises beyond the linear response approximation
in that the operators corresponding to various properties of
quarkonium states need to be evaluated in an ensemble of anisotropic
(in momentum space) gauge field configurations.

After reviewing the anisotropic plasma in Section \ref{sec-aniso}, we
formulate the first potential model for an anisotropic medium in
Section \ref{section3}. The solution of the 3-dimensional
Schr\"odinger equation is described in section~\ref{SEq3d}, and the
numerical results are presented in section~\ref{sec:Results}. We note
that by now it is understood that merely solving the Schr\"odinger
equation for the individual states is not enough, especially close to
the continuum threshold. Here many-body interactions should be taken
into consideration by solving the Schr\"odinger equation for the
non-relativistic Green's function \cite{Mocsy:2007yj}. However, for
the purpose of the present work the simple analysis suffices. We
summarize our results and draw our conclusions in Section
\ref{sec:Summary}.

\section{The anisotropic plasma}
\label{sec-aniso}

The phase-space distribution of gluons is assumed to be given by the
following {\em ansatz}~\cite{Dumitru:2007hy,Romatschke:2003ms,
Mrowczynski:2004kv,Romatschke:2004jh,Schenke:2006fz}:
\begin{equation}
f({\bf p}) = f_{\rm iso}\left(\sqrt{{\bf p}^2+\xi({\bf p}\cdot{\bf
n})^2} \right) ~.  \label{eq:f_aniso}
\end{equation}
Thus, $f({\bf p})$ is obtained from an isotropic distribution $f_{\rm
  iso}(|{\bf{p}}|)$ by removing particles with a large momentum
component along ${\bf{n}}~$, the direction of anisotropy. We shall
restrict ourselves here to a plasma close to equilibrium, which is
motivated by the fact that in a heavy-ion collision quarkonium states
are expected to form when the temperature has dropped to (1-2)~$T_c$;
by then, the plasma may have equilibrated at least partly. Hence, we
assume that the function $f_{\rm iso}(|{\bf{p}}|)$ is a thermal
ideal-gas distribution.

The parameter $\xi$ determines the degree of anisotropy,
\beq
\xi = \frac{1}{2} \frac{\langle {\bf p}_\perp^2\rangle}
{\langle p_z^2\rangle} -1~,
\eeq
where $p_z\equiv \bf{p\cdot n}$ and ${\bf p}_\perp\equiv {\bf{p-n
(p\cdot n)}}$ denote the particle momentum along and perpendicular to
the direction ${\bf n}$ of anisotropy, respectively. If $\xi$ is small
then it is also related to the shear viscosity of the plasma; for
example, for one-dimensional boost-invariant expansion~\cite{Asakawa:2006tc}
\beq \label{eq:xi_eta}
\xi = \frac{10}{T\tau} \frac{\eta}{s}~,
\eeq
where $T$ is the temperature, $\tau$ is proper time (and $1/\tau$ is
the Hubble expansion rate), and $\eta/s$ is
the ratio of shear viscosity to entropy density. In an expanding
system, non-vanishing viscosity (finite momentum relaxation
rate) implies an anisotropy of the particle momenta which
increases with the expansion rate $1/\tau$. For $\eta/s\simeq
0.1$ -- 0.2 and $\tau T\simeq1$ -- 3 one finds that $\xi\simeq1$.

We should stress that in this paper we restrict to solving the
time-independent Schr\"odinger equation, i.e.\ we assume that the
plasma is at a constant temperature $T$ and anisotropy $\xi$. This
approximation is useful if the time scale associated with the bound
state, $\sim 1/|E_{\text{bind}}|$, is short compared to the time
scales over which $T$ and $\xi$ vary. Indeed, for sufficiently large
quark mass $m_Q$ this condition should be satisfied.

\section{The Karsch-Mehr-Satz model at finite temperature}
\label{section3}

\subsection{Isotropic medium ($\xi=0$)}

Lacking knowledge of the exact heavy-quark potential at finite
temperature, different phenomenological potentials, as well as
lattice-QCD based potentials have been used in potential models to
study quarkonium.

The KMS model~\cite{KMS} assumes the following form of the heavy-quark
potential at finite temperature in an isotropic plasma with $\xi=0$: 
\beq \label{KMS_free}
F(r,T) = -\frac{\alpha}{r} \exp\left( -m_D
\, r  \right) + \frac{\sigma}{m_D}\left[1-\exp\left( -m_D
\, r  \right)\right]\, .
\eeq
Here, $\alpha\approx 0.385$ is an effective Coulomb coupling at
(moderately) short distances, $\sigma=0.223$~GeV$^2$ is the string
tension and $m_D(T)$ is the Debye screening mass.

Eq.~(\ref{KMS_free}) is a model for the action of a Wilson loop of
size $1/T$ and $r$ in the temporal and spatial directions,
respectively (see~\cite{Petreczky:2005bd} and references therein).
This potential has been used before to study quarkonium bound states
\cite{mocsy}. However, it was realized shortly after that
eq.~(\ref{KMS_free}) cannot be taken directly as the heavy-quark
potential because it contains an entropy contribution; see, for
example, the discussion in
refs.~\cite{Mocsy:2007yj,Shuryak:2004tx,Petreczky:2005bd,Mocsy:2008eg}.
Rather, eq.~(\ref{KMS_free}) corresponds to the free energy due to the
presence of a $Q\overline{Q}$ in the medium. We emphasize that the
entropy term in the {\em lattice data} is merely a perturbative
entropy contribution present at large distances\footnote{We evaluate
  it in anisotropic HTL resummed perturbation theory in
  section~\ref{sec:pertSelf}.}, $r\rightarrow\infty$, and it is absent at
short distances~\cite{Petreczky:2005bd}. One can remove this entropy
term from the lattice data by parameterizing
$F(r\rightarrow\infty,T)\equiv F_\infty(T)$ in the form
$F_\infty(T)=a/T-bT$ and then adding the term $bT$ to $F(r,T)$ at
large distance, thereby obtaining what has been called {\it the
  physical potential} in \cite{Mocsy:2007yj,Mocsy:2007jz}.

Alternatively, one could calculate the full entropy $S=- \partial
F/\partial T$ and add it to the free energy, which leads to the
internal energy $U=F+TS$. The internal energy calculated in lattice
QCD \cite{kaczmarekHP} shows a large increase in $U_\infty$ near
$T_c$, due to the large increase of the entropy near
$T_c$. Furthermore, at temperatures $T\simeq T_c$ a potential model
based on the internal energy becomes much more binding than the $T=0$
Cornell potential (we refer to this as "overshooting"). For these
reasons, the internal energy $U(r,T)$ obtained on the lattice should
neither be identified with the heavy quark potential, although it has
been used in potential models
before~\cite{Shuryak:2004tx,rapp,mocsy}. Nevertheless, the internal
energy provides a useful upper limit for the potential at finite $T$.
A version of the internal energy in which the overshooting problem was
eliminated, was designed in \cite{Mocsy:2007yj,Mocsy:2007jz} and
called the {\it most confining potential}.

\begin{figure}
\includegraphics[width=10cm]{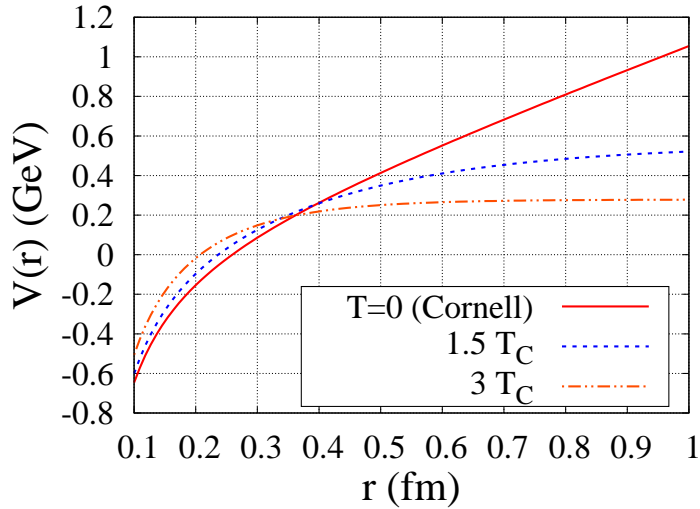}
\caption[a]{The model potential from eq.~(\ref{KMSpot}) at zero and at
finite temperature as a function of distance. Temperature is
normalized to $T_c=192$~MeV and the temperature dependence of the
Debye mass is parameterized as given in eq.~(\ref{eq:mD(T)}) below.}
\label{fig:modKMS}
\end{figure}

In this paper we also construct a model for a potential which could be
viewed as an upper limit for the heavy quark potential,
i.e.\ $V_\infty\simeq U_\infty$. Our present model is very simple and
contains a minimum number of parameters, as the primary goal is to
generalize the finite-$T$ potential to anisotropic media. In our model
we add the full entropy contribution to the
KMS {\em ansatz}~(\ref{KMS_free}):
\begin{eqnarray}
V(r,T) &=& F(r,T) - T\frac{\partial F(r,T)}{\partial T} \label{eq:KMSpotFE}\\
&\approx& -\frac{\alpha}{r} \left(1+m_D \, r\right) 
\exp\left( -m_D \, r  \right)
+ 2 \frac{\sigma}{m_D}\left[1-\exp\left( -m_D \, r  \right)\right]
- \sigma \,r\, \exp(-m_D\,r)\, . \label{KMSpot}
\end{eqnarray}
In the second line we have used that $m_D$ is approximately
proportional to $T$ at high temperatures. Since the effect of the
running of the coupling is important only at distances $<0.1~$fm, not
relevant for quarkonium studies, here we do not consider
running-coupling corrections. Fig.~\ref{fig:modKMS} compares the
potential at finite temperature to that at $m_D=0$ which is a Cornell
potential.

This potential, just as its original form~(\ref{KMS_free}),
essentially represents an interpolation from the well-known Cornell
potential at short distance to an exponentially Debye-screened string
attraction at large $r$. With $g\simeq2$, $m_D\simeq gT$ and
$T_c\simeq200$~MeV, the length scale where medium effects become large
is roughly given by $r_{\rm med}(T) \simeq T_c/(2T)~$fm, in
approximate agreement with lattice results from
ref.~\cite{Kaczmarek:2004gv}. In~(\ref{KMSpot}) corrections to the
Cornell potential are suppressed at distances $r<1/m_D$, i.e.\ they
appear only at order $(m_D\, r)^2$. This is due to the fact that we
subtracted the derivative $\partial F/\partial \log T$ even at
intermediate distances; it appears to give a better representation of
the lattice potential at $r<r_{\rm med}(T)$, which in fact coincides
with the Cornell {\em ansatz}. One can see in Fig.~\ref{fig:modKMS}
that our potential $V(r)$ is very close to the Cornell potential for distances
up to $r\simeq0.4$~fm, in agreement with lattice
results~\cite{Kaczmarek:2004gv}.  The finite-temperature potential
(\ref{KMSpot}) does not overshoot the Cornell potential significantly
at any $r$~\cite{Mocsy:2007jz}, at least up to temperatures on the
order of $1.5\, T_c$. This is actually the temperature range where
most bound states (except perhaps 1S bottomonium) are expected to
dissociate in an isotropic medium \cite{Mocsy:2007yj,Mocsy:2007jz}. On
the other hand, Fig.~\ref{fig:modKMS} shows that at rather high
temperatures of order $3T_c$, the model~(\ref{KMSpot}) {\em does}
overshoot the Cornell potential at short distances. This indicates
that this simple form of the finite-$T$ potential is not appropriate
when the Debye mass $m_D$ is large. However, this regime is not of
interest here since even the $b\overline{b}$ states are no longer
bound. Overall, the potential~(\ref{KMSpot}) appears to provide a
reasonable model for the inter-quark potential in the deconfined phase
at (moderately) high temperatures.

At $r\to\infty$ the potential~(\ref{KMSpot}) approaches
\beq \label{eq:modKMS_Vinf}
V_\infty(T) = 2\frac{\sigma}{m_D} \simeq \frac{0.16~{\rm GeV}^2}{T}~.
\eeq
Again, this is in approximate agreement with the $V_\infty\simeq 1/T$
{\rm ansatz} used in ref.~\cite{Mocsy:2007jz}. Note, in particular,
that~(\ref{eq:modKMS_Vinf}) is about the same as the internal energy
$U_\infty(T)$ obtained from the lattice data~\cite{Mocsy:2007jz}. We
take this as an indication that our potential (\ref{KMSpot})
represents an upper limit for the possible finite temperature
potentials.

The main assumption of the KMS model is that the very same screening
scale $m_D$ which emerges in the Debye-Coulomb potential also appears
in the non-perturbative, long-distance contribution due to the string.
In the following, we take over this assumption to anisotropic plasmas.

It is interesting to note that the KMS {\em ansatz} for the free
energy from eq.~(\ref{KMS_free}) can be obtained in the usual way
from the Fourier transform of the static propagator, provided that a
non-perturbative contribution 
\beq \label{NP_prop} 
\frac{m_G^2}{({\bf k}^2+m_D^2)^2} 
\eeq 
is added to the standard HTL resummed
propagator~\cite{Megias:2007pq}. Here, $m_G^2$ is a constant of
dimension two which can be related to the string tension $\sigma$ by
matching onto a Cornell potential at small $m_D\,r$. The presence of
such an additional dimensionful scale (besides $T$) also leads to a
non-vanishing trace of the energy-momentum tensor~\cite{Megias:2009mp}.

\subsection{Anisotropic medium ($\xi>0$)} \label{sec:anisoKMS}

Our fundamental assumption is that the modified KMS potential~(\ref{KMSpot}),
which provides a reasonable upper-limit model for the heavy-quark potential in
isotropic media, retains its basic form also when the local momentum
distribution of the plasma particles is anisotropic
($\xi>0$). However, the isotropic Debye mass $m_D(T)$ is now replaced
by an angular dependent screening scale $\mu(\theta;\xi,T)$ as
discussed in the next section.

\subsubsection{Angular dependence of the potential at short distances}

The potential from one-gluon exchange at short distance can be
evaluated perturbatively. At tree level the potential corresponds to
the Fourier transform of the static gluon propagator which resums
screening effects at high temperature\footnote{In the first part of
  this section we are only interested in distances up to $\sim
  1/\mu$. It is therefore not crucial to distinguish carefully whether
  the potential is identified with the Fourier transform of the
  propagator or with the internal energy. We shall add the entropy
  contribution later, c.f.\ eqs.~(\ref{KMSpot_xi}-\ref{KMSpot_xi2}),
  to suppress corrections to the Cornell potential at short
  distances.}. For $\xi>0$ the potential depends not only on the
distance between the $Q$ and $\overline{Q}$ but also on the angle
$\theta$ between their separation ${\bf r}$ and the direction ${\bf
  n}$ of anisotropy~\cite{Dumitru:2007hy}.

To linear order in $\xi$, the potential can be expressed as
\bea \label{eq:anisoPotlin_xi}
V({\bf{r}},\xi) &=& V_{\rm iso}(r)-g^2 C_F \, \xi m_D^2 \int
\frac{d^3{\bf{p}}}{(2 \pi)^3} \, e^{i{\bf{p \cdot r}}} \,
\frac{\frac{2}{3}-({\bf p\cdot
n})^2/{\bf{p}}^2}{({\bf{p}}^2+m_D^2)^2} \\
&=& {V}_{\rm iso}(r) \left(1-\xi {\cal F}(\hat{r},\theta) \right)~,
\eea
where $\hat{r}\equiv r\,m_D$ and $m_D(T)$ denotes the screening mass
in the isotropic medium at a given temperature $T$, as
before. Also, ${V}_{\rm iso}(r)$ is the Debye-screened Coulomb potential
in an isotropic medium, as given by the first term in
eq.~(\ref{KMS_free}), and the function
\beq
{\cal F}(\hat{r},\theta) \equiv f_0(\hat{r})+f_1(\hat{r})\cos(2\theta)
\eeq
with
\begin{eqnarray}
f_0(\hat{r}) &=&
  \frac{6(1-e^{\hat{r}})+\hat{r}[6-\hat{r}(\hat{r}-3)]}{12\hat{r}^2}
= -\frac{\hat{r}}{6}-\frac{\hat{r}^2}{48}+\cdots~,\label{eq:f0}\\
f_1(\hat{r}) &=&
  \frac{6(1-e^{\hat{r}})+\hat{r}[6+\hat{r}(\hat{r}+3)]}{4\hat{r}^2}
= -\frac{\hat{r}^2}{16}+\cdots~.
\label{eq:f1}
\end{eqnarray}
Note that eqs.~(\ref{eq:anisoPotlin_xi}-\ref{eq:f1}) do not apply at
large distances $\hat{r}\gg1$, which is a shortcoming of the Taylor
expansion of the full potential in powers of $\xi$. This is of no
importance in the following because these expressions are used at
short distances only in order to determine the angular dependence of
the screening scale, which will then replace $m_D$ in
eq.~(\ref{KMSpot}).

We can now define the $\theta$-dependent screening mass in the
anisotropic medium as the inverse of the distance scale $r_{\rm
med}(\theta)$ over which $|rV(r)|$ drops by a factor of $e$:
\bea
\log \frac{-\alpha}{{r}_{\rm med} \, V(r_{\rm med},\theta;\xi,T)}
&=& 1~, \label{eq:logV_rmed}\\
\mu(\theta;\xi,T) &=& {r_{\rm med}^{-1} (\theta;\xi,T)}~. 
\eea
In~(\ref{eq:logV_rmed}) we have used the fact that $rV\to-\alpha$ as
$r\to0$. To leading order in $\xi$ this leads to
\begin{eqnarray}
\hat{r}_{\rm med} &=& 1-\xi {\cal F}(\hat{r}_{\rm
med},\theta) \label{eq:rmed_xi}~.
\end{eqnarray}
An approximate solution to eq.~(\ref{eq:rmed_xi}) is given by
\beq \label{rmed_approx} 
\hat{r}_{\rm med} \simeq 1+\xi \frac{3+\cos
2\theta}{16}~~~\Rightarrow~~~ \frac{\mu}{m_D} \simeq 1-\xi
\frac{3+\cos 2\theta}{16}~. \label{mu}
\eeq
For $\xi=0.5$ (1.0) this solution achieves a relative accuracy of
$\le4\%$ ($\le18\%$) over the entire range of $\theta$.  The accuracy
of this result can be improved systematically by going beyond ${\cal
O}(\hat{r}^2)$ in the expansion of the functions $f_0$ and $f_1$
introduced in eqs.~(\ref{eq:f0},\ref{eq:f1}).

The ``minimal'' extension of the KMS model to non-zero anisotropy
consists of replacing $m_D(T)$ in~(\ref{KMSpot}) by
$\mu(\theta;\xi,T)$ from above:
\beq \label{KMSpot_xi}
V({\bf r}) = -\frac{\alpha}{r} \left(1+\mu \, r\right) \exp\left( -\mu
\, r  \right) + \frac{2\sigma}{\mu}\left[1-\exp\left( -\mu
\, r  \right)\right]
- \sigma \,r\, \exp(-\mu\,r)~.
\eeq
At short distances this reduces to the Cornell potential $V(r) =
-{\alpha}/{r} +\sigma\,r$, as it should be.

Corrections to eq.~(\ref{KMSpot_xi}) due to the finite quark mass can
be accounted for by adding a temperature- and spin-independent
correction proportional to $\sigma/(m_Q^2\, r)$~\cite{Bali}. This
improves the accuracy of the wave functions of quarkonium states
obtained from the solution of the non-relativistic Schr\"odinger
equation. The potential finally takes the form
\beq \label{KMSpot_xi2}
V({\bf r}) = -\frac{\alpha}{r} \left(1+\mu \, r\right) \exp\left( -\mu
\, r  \right) + \frac{2\sigma}{\mu}\left[1-\exp\left( -\mu
\, r  \right)\right]
- \sigma \,r\, \exp(-\mu\,r)- \frac{0.8 \, \sigma}{m_Q^2\, r}~.
\eeq
%

\subsubsection{Perturbative heavy-quark free energy in an anisotropic
  medium} \label{sec:pertSelf}
  
In the limit of infinite mass, the free energy of a heavy quark in a
thermal plasma is related to the expectation value of Polyakov
loops. At high temperature, this can be calculated within hard-thermal
loop resummed perturbation theory. The leading-order contribution is
given by
\beq \label{eq:SelfE}
F_Q(\xi,T) = - \frac{1}{2}
(ig)^2 \, C_F \int \frac{d^3k}{(2\pi)^3} \left[ \Delta^{00}({\bf k}) -
\frac{1}{{\bf k}^2} \right] ~,
\eeq
where $\Delta^{00}({\bf k})$ is the ``hard thermal loop'' resummed
propagator of static $A_0$ fields. We have subtracted the
temperature-independent contribution to $F_Q$ which is not of interest
here. Also, this renders the integral UV-finite. For the case of an
anisotropic medium, $\Delta^{00}({\bf k})$ has been calculated in
ref.~\cite{Dumitru:2007hy}; see, also, ref.~\cite{MTB}. The
expression~(\ref{eq:SelfE}) then turns into
\beq \label{eq:SelfExi}
F_Q(\xi,T) =  - \frac{1}{2} \alpha_s C_F m_D h(\xi)~,
\eeq
with a temperature-independent function $h(\xi)$. For small anisotropy
the following expansion in $\xi$ applies:
\beq
h(\xi) = 1 - \frac{1}{6}\xi + \frac{18-\pi^2}{240} \xi^2 + \cdots
\eeq
We identify this expression with the free energy of a quark due to its
interaction with the medium (i.e., half the free energy of a
$Q\overline{Q}$ pair at large separation). However, the
perturbative entropy contribution $TS=-T\partial F_Q/\partial T$
should be added again in order to obtain the potential.
Since~(\ref{eq:SelfExi}) is linear in $T$ (at fixed coupling), it
follows that the perturbative contribution to $V_\infty$
vanishes\footnote{A term $F_Q\sim a/T$ could be
  generated by a non-perturbative contribution of the form
  $m_G^2/({\bf k}^2 + m_D^2)^2$ to the static gluon propagator, as
  already mentioned above. $m_G^2$ is a constant of dimension
  two~\cite{Megias:2007pq}.}. On the other hand, lattice data (for an
  isotropic medium) indicate that the free energy of a quark-antiquark
  pair at infinite separation also contains a non-perturbative
  contribution of the form $V_\infty(T)=r_{\rm med}(T) \,
  \sigma$~\cite{Mocsy:2007yj}, which agrees qualitatively with the
  prediction of the KMS model, $V_\infty \sim {\sigma}/m_D$. Within
  the framework of the KMS model, this implies that at $\xi\neq0$,
  $V_\infty(\theta;\xi,T)$ depends on angle as screening becomes
  anisotropic.

\section{Solving the 3d Schr\"odinger Equation} \label{SEq3d}

To determine the wave functions of bound quarkonium states, we solve
the Schr\"odinger equation
\bea
\hat{H} \phi_\upsilon({\bf x}) &=& E_\upsilon \, \phi_\upsilon({\bf
x})  ~, \nonumber \\
\hat{H} &=& -\frac{\nabla^2}{2 m_R} + V({\bf x}) + m_1 + m_2~,
\label{3dSchrodingerEQ}
\eea
on a three-dimensional lattice in coordinate space with the potential
given in eq.~(\ref{KMSpot_xi2}).  Here $m_1$ and $m_2$ are the masses
of the two heavy quarks and $m_R$ is the reduced mass: $m_R = m_1
m_2/(m_1+m_2)$. The index $\upsilon$ on the eigenfunctions,
$\phi_\upsilon$, and energies, $E_\upsilon$, represents a list of all
relevant quantum numbers, e.g.\ $n$, $l$, and $m$ for a radial Coloumb
potential. Due to the anisotropic screening scale, the wave functions
are no longer radially symmetric if $\xi \neq 0$. Since we consider
only small anisotropies we nevertheless label the states as $1S$
(ground state) and $1P$ (first excited state), respectively.

To find solutions to Eq.~(\ref{3dSchrodingerEQ}) we use the finite
difference time domain method (FDTD)~\cite{Sudiarta:2007}.  In this
method we start with the time-dependent Schr\"odinger equation
\beq
i \frac{\partial}{\partial t} \psi({\bf x},t) = \hat H \psi({\bf x},t) \, ,
\label{3dSchrodingerEQminkowski}
\eeq
which can be solved by expanding in terms of the eigenfunctions,
$\phi_\upsilon$:
\beq \psi({\bf x},t) = \sum_\upsilon c_\upsilon \phi_\upsilon({\bf x})
e^{- i E_\upsilon t}~.
\label{eigenfunctionExpansionMinkowski}
\eeq
If one is only interested in the lowest energy states (ground state
and first few excited states) an efficient way to proceed is to
transform~(\ref{3dSchrodingerEQminkowski})
and~(\ref{eigenfunctionExpansionMinkowski}) to Euclidean time by a
Wick rotation, $\tau \equiv i t$:
\beq \frac{\partial}{\partial \tau} \psi({\bf x},\tau) = - \hat H
\psi({\bf x},\tau) \, ,
\label{3dSchrodingerEQeuclidean}
\eeq
and
\beq \psi({\bf x},\tau) = \sum_\upsilon c_\upsilon \phi_\upsilon({\bf
x}) e^{- E_\upsilon \tau} ~.
\label{eigenfunctionExpansionEuclidean}
\eeq
For details of the algorithm we refer to ref.~\cite{Sudiarta:2007}.

\subsection{Finding the ground state}

By definition the ground state is the state with the lowest energy
eigenvalue, $E_0$. Therefore, at late imaginary time the sum
over eigenfunctions (\ref{eigenfunctionExpansionEuclidean}) is
dominated by the ground state eigenfunction
\beq \lim_{\tau \rightarrow \infty} \psi({\bf x},\tau) \rightarrow c_0
\phi_0({\bf x}) e^{- E_0 \tau}~.
\label{groundstateEuclideanLateTime}
\eeq
Because of this one can obtain the ground state wavefunction,
$\phi_0$, and energy, $E_0$, by solving
Eq.~(\ref{3dSchrodingerEQeuclidean}) starting from a random
three-dimensional wavefunction, $\psi_{\text{initial}}({\bf x},0)$,
and evolving forward in imaginary time. This initial wavefunction
should have nonzero overlap with all eigenfunctions of the
Hamiltonian; however, due to the damping of higher-energy
eigenfunctions at sufficiently late imaginary times we are left with
only the ground state, $\phi_0({\bf x})$. Once the ground state
wavefunction (or, in fact, any other wavefunction) is found we can
compute its energy eigenvalue via
\bea
E_\upsilon(\tau\to\infty) = \frac{\langle \phi_\upsilon | \hat{H} |
\phi_\upsilon \rangle}{\langle \phi_\upsilon | \phi_\upsilon
\rangle} = \frac{\int d^3{\bf x} \, \phi_\upsilon^*
\, \hat{H} \, \phi_\upsilon }{\int d^3{\bf x} \, \phi_\upsilon^*
\phi_\upsilon} \; .
\label{bsenergy}
\eea

To obtain the binding energy of a state,
$E_{\upsilon,\text{bind}}$, we subtract the quark masses and
the potential at infinity
\beq
E_{\upsilon,\text{bind}} \equiv E_\upsilon - m_1 - m_2 -
\frac{\langle \phi_\upsilon | V(\theta,|{\bf r}|\to\infty) | \phi_\upsilon
\rangle}{\langle \phi_\upsilon | \phi_\upsilon \rangle} \; .
\label{bsbindingenergy}
\eeq
For the isotropic KMS potential the last term is independent of the
quantum numbers $\upsilon$ and equal to $\sigma/m_D$. In the
anisotropic case, however, this is no longer true since the operator
$V_\infty(\theta)$ carries angular dependence, as already discussed
above. Its expectation value is of course independent of $\theta$ but
does depend on the anisotropy parameter $\xi$.

\subsection{Finding the excited states}

The basic method for finding excited states is to first evolve the
initially random wavefunction to large imaginary times, find the
ground state wavefunction, $\phi_0$, and then project this state out
from the initial wavefunction and re-evolve the partial-differential
equation in imaginary time. However, there are (at least) two more
efficient ways to accomplish this. The first is to record snapshots of
the 3d wavefunction at a specified interval $\tau_{\text{snapshot}}$
during a single evolution in $\tau$. After having obtained the ground
state wavefunction, one can then go back and extract the excited
states by projecting out the ground state wavefunction from the
recorded snapshots of $\psi({\bf x},\tau)$.

An alternative way to select different excited states is to impose a
symmetry condition on the initially random wavefunction which cannot
be broken by the Hamiltonian evolution. For example, one can select
the first excited state of the (anisotropic) potential by
anti-symmetrizing the initial wavefunction around either the $x$, $y$,
or $z$ axes.  In the anisotropic case this trick can be used to
separate the different polarizations of the first excited state of the
quarkonium system and to determine their energy eigenvalues with high
precision.  This high precision allows one to more accurately
determine the splitting between polarization states which are
otherwise degenerate in the isotropic Debye-Coulomb potential.

Whichever method is used, once the wave function of an excited state
has been determined one can again use the general
formulas~(\ref{bsenergy}) and~(\ref{bsbindingenergy}) to determine its
binding energy.

\section{Results and Discussion} 
\label{sec:Results}

In this section we present the solutions of the 3-dimensional
Schr\"odinger equation (\ref{3dSchrodingerEQ}) in a weakly anisotropic
medium. In particular, we determine the temperature dependence of the
binding energies of different charmonium and bottomonium states
obtained with the anisotropic potential~(\ref{KMSpot_xi2}) that has
been contructed from the {\it most binding} isotropic potential. The
anisotropy- and temperature-dependent screening mass
$\mu(\theta;\xi,T) $ is given in equation (\ref{mu}). To illustrate
the effect of the anisotropy of the medium more clearly we shall also
compare the results to those obtained for an isotropic medium. In the
latter case $\xi=0$ and so $\mu(\theta;\xi,T) =m_D(T)$, where the
temperature dependence of the Debye mass is given by
\beq \label{eq:mD(T)}
m_D(T) = A \, gT\, \sqrt{(1+{n_f}/{6})} \, .
\eeq
For $n_f=2$ number of massless quark flavors the parameter $A=1.4$ has
been determined in lattice calculations \cite{Kaczmarek:2005}. We
choose a fixed gauge coupling of $g=1.72$ which yields
$m_D(T)/T\approx2.8$. This agrees approximately with lattice estimates
of $m_D/T$ for temperatures on the order of $T/T_c\sim1.5$, and it
also gives a reasonable estimate of the free energy at infinite
separation~\cite{Kaczmarek:2005}. The values of the charm and bottom
quark masses are chosen such that at low temperature the correct
masses of $M_{J/\psi}=3.1~$GeV and $M_{\Upsilon}=9.4~$GeV for the
$J/\psi$ and the $\Upsilon$, respectively, are recovered. Accordingly,
\beq
m_c = 1.3~{\rm GeV} ~~~~~ \mbox{and} ~~~~~ m_b=4.7~{\rm GeV}\, .
\eeq

All of the results reported below were obtained from lattices with
lattice spacings approximately 20 times smaller than the
root-mean-square radius ${\langle {\bf r}^2\rangle^{1/2}_\upsilon(T,
 \xi, m_Q)}$ of the state under consideration, defined by the quantum
number $\upsilon$. The lattice size $L$ was chosen to be about 6 times
larger than the RMS radius\footnote{Since we restrict the analysis to
 only weak medium anisotropies, we employ isotropic lattices with
 uniform lattice spacing in all three cartesian
 directions.}. Discretization errors and finite-size effects are thus
expected to be reasonably small and nearly independent of $T$, $\xi$,
$m_Q$, and $\upsilon$. We stopped the time evolution when the energy
$E_\upsilon(\tau)$ of the state had stabilized to within $10^{-8}$. A
more detailed investigation of numerical errors is beyond the scope of
the present paper. Our goal here is to show how quarkonium states may
be affected by the anisotropy of the medium.

\begin{figure}
\includegraphics[width=14cm]{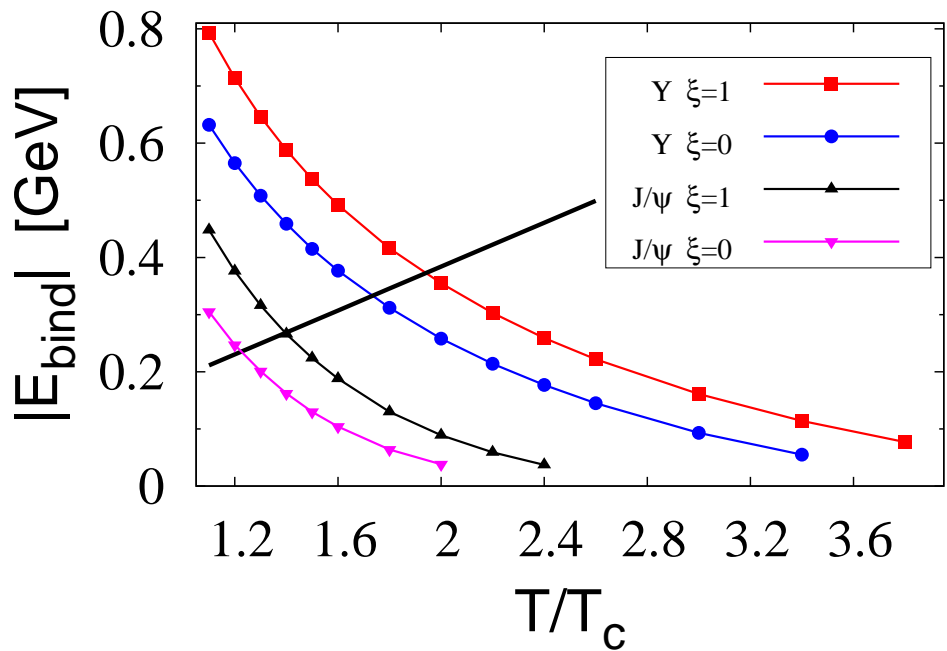}
\vspace*{-1cm}
\caption[a]{Temperature-dependence of the binding energies
 $|E_{\text{bind}}|$ for the ground-states of charmonium (lower
 curves) and bottomonium (upper curves) in the vector channel for two
 values of the plasma anisotropy parameter $\xi$. The straight line
 corresponds to a binding energy equal to the temperature.}
\label{fig:Eb_bott}
\end{figure}
\begin{figure}
\includegraphics[width=14cm]{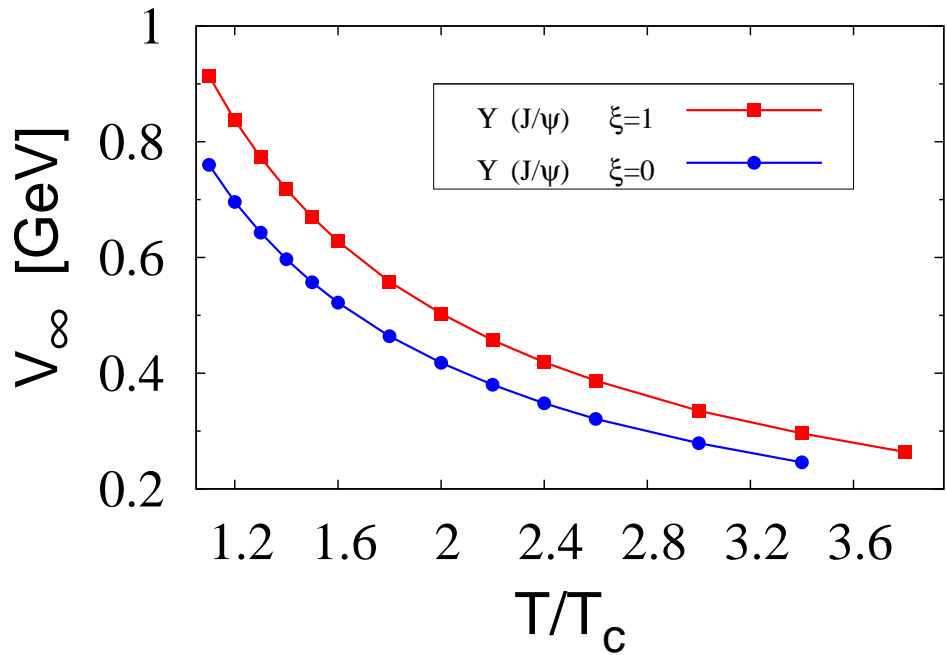}
\vspace*{-1cm}
\caption[a]{Expectation value of $V_\infty$ in the $J/\Psi$ or
$\Upsilon$ states as a
function of temperature for two values of the plasma anisotropy
parameter $\xi$.}
\label{fig:Vinf}
\end{figure}

The temperature dependence of the binding energies of charmonium and
bottomonium ground states in the vector channel\footnote{Spin effects
 are neglected in our treatment and the ground state could be
 identified with either the pseudo-scalar or the vector state. For
 definitness, we shall refer to the vector channel.} are depicted
in Fig.~\ref{fig:Eb_bott}. The figure shows the results obtained for
isotropic $\xi=0$ and anisotropic $\xi=1$ media. The former are in agreement with those obtained in ref.~\cite{Mocsy:2007jz} with the so-called {\it most-confining} isotropic potential. As expected, the
binding energy decreases as the screening mass $m_D(T)$ increases with
temperature $T$. This plot also indicates that $|E_{\text{bind}}|$
increases with the anisotropy $\xi$. This can be understood
from the fact that in an anisotropic plasma the screening scale
$\mu(\theta)$ at a given temperature is smaller than the corresponding
Debye mass $m_D$; see eq.~(\ref{rmed_approx}).
As a consequence, the screening of the attractive Coulomb and string
contributions is less accentuated in the anisotropic plasma and so
quarkonium states are bound more strongly than in an isotropic
medium. The magnitude of this effect is substantial even for the
moderate anisotropy considered here. Near the critical
temperature $T_c$, for example, the binding energy of the 1S vector
$c\bar{c}$ ground state increases by about $50\%$, and that of the 1S
$b\bar{b}$ ground state increases by roughly $30\%$ compared to
the binding energies calculated in an isotropic medium (the only case
addressed previously in the literature).

It is important to highlight another aspect of the reduced
screening. In the KMS model the asymptotic value of the potential is
intrinsically related to the screening mass via the relation $\langle
V_\infty\rangle(T)=\langle\upsilon| \sigma \mu^{-1}(\theta;\xi,T) |
\upsilon\rangle$.  This implies that in the anisotropic medium less
screening translates into an increase of the potential at infinite
separation, $V_\infty$. The above is illustrated in
Fig.~\ref{fig:Vinf} which shows the expectation value of $V_\infty$ in
the $\Upsilon$ state\footnote{We recall that $V_\infty$ is
  proportional to the identity at $\xi=0$ and hence its expectation
  value is the same for all states. At $\xi=1$ we obtain a very small
  difference between $\langle J/\Psi | V_\infty | J/\Psi \rangle$ and
  $\langle \Upsilon | V_\infty | \Upsilon \rangle$.}. $V_\infty$, in
turn, determines the continuum threshold, which, at a given $T$ is at
higher energy than in the isotropic case. This implies that at a given
temperature the energy gap between the bound state and the continuum,
which is the binding energy, is increased compared to the isotropic
case.

Comparing the behavior of $\langle V_\infty\rangle$ to that of the
binding energy of the $\Upsilon$ from fig.~\ref{fig:Eb_bott} shows
that the decrease of $|E_{\text{bind}}|$ with $T$ is largely due to
the decrease of the continuum threshold $\langle V_\infty
\rangle$. Indeed, we have confirmed that the wave function of the
$b\bar{b}$ ground state is essentially unaffected by the medium, i.e.\
it is almost independent of $T$ (for $T\simle 2T_c$) and $\xi$ (for
$\xi\simle1$). This has interesting implications for phenomenology: On
one hand, the center of the $\Upsilon$-peak in the dilepton invariant
mass distribution may not shift much (since $V_\infty=0$ for decay
into dileptons) even for temperatures $> T_c$ where the binding energy
is significantly lower than in vacuum. On the other hand, when
$|E_{\text{bind}}|\sim T$ we expect substantial broadening of the
states due to direct thermal activation \cite{dima,Mocsy:2007jz}. The
thermal width can be estimated from the binding
energy~\cite{dima}. When the width is larger than the binding energy,
a state decays faster than it binds \cite{Mocsy:2007jz}. Note, that in
the same temperature domain collisions with thermal particles of the
medium would further broaden the width of a state. Thus, the
dissociation of the bound states may be expected to occur roughly when
$|E_{\text{bind}}|\sim T$~\cite{Mocsy:2008eg}. With the potential
investigated in this paper, which likely represents an upper limit for
the attractive interaction, the condition $|E_{\text{bind}}|\sim T$ is
met for the $J/\psi$ by $1.2\, T_c$ for $\xi=0$, in agreement with
previous results \cite{Mocsy:2007jz}, and by $1.4\, T_c$ for
$\xi=1$. We stress furthermore that the thermal density of a given
state,
\beq \rho \sim
\exp\left(-\frac{E_{\text{bind}}}{T}\right)  \label{eq:Boltz}\, ,
\eeq 
is not enhanced significantly when $|E_{\text{bind}}|< T$. In other
words, since $T$ decreases with time in a heavy-ion collision,
quarkonium states with quantum numbers $\upsilon$ should appear at a
temperature $T_\upsilon \sim |E_{\upsilon,\text{bind}}|$ \footnote{The
  total number of formed quarkonium states depends on how many heavy
  quarks are produced in the initial hard processes, and on what
  fraction thereof is bound in $D$- and $B$-mesons,
  respectively.}. From fig.~\ref{fig:Eb_bott} it is plausible that in
a viscous plasma quarkonium synthesis occurs at higher temperature
than in a perfectly equilibrated medium. For the $J/\Psi$ for example,
$\Delta T_{\text{synth}}/T_c \simeq 20\%$ for $\xi=1$ as compared to
$\xi=0$.

\begin{figure}
\includegraphics[width=14cm]{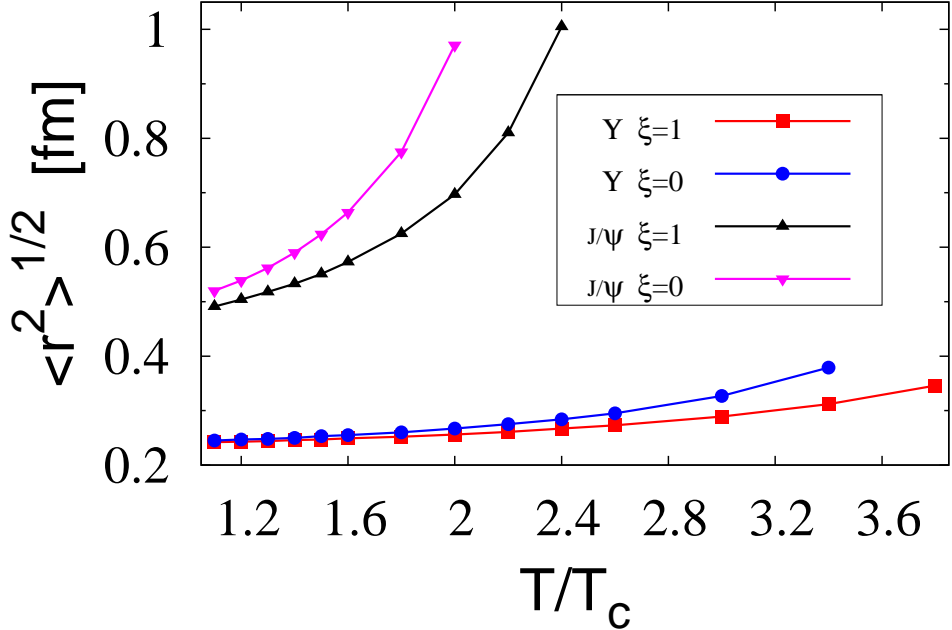}
\vspace*{-1cm}
\caption[a]{Root-mean-square radii of the $1S$ ground-states of
charmonium (upper curves) and bottomonium (lower curves) as
functions of temperature for two values of the plasma anisotropy
parameter $\xi$.}
\label{fig:rRMS}
\end{figure}
In fig.~\ref{fig:rRMS} we show the root-mean-square (RMS) radii
${\langle {\bf r}^2\rangle^{1/2}(T, \xi)}$ of the $c\bar{c}$ and
$b\bar{b}$ ground states as functions of temperature. The former grows
rather rapidly about the dissociation point where
$|E_{\text{bind}}|\sim T$.  The size of the $\Upsilon$, on the other
hand, increases only little with temperature. We can understand these
results, qualitatively, as follows. For charmonium the string part of
the potential dominates, and the growth of its RMS radius with $T$
indicates that screening of the string is strong\footnote{Recall that
  the in the KMS model the string tension enters with a factor of
  $1/m_D(T)$ at intermediate distances on the order of
  $r\sim1/m_D(T)$.}. We observe a similar behavior of the first $1P$
excited state of bottomonium. On the other hand, $1S$ bottomonium is
too small to be affected strongly by screening (for $T\simle 2T_c$),
it is essentially a Coulomb state. The weaker binding as
compared to low temperature is largely due to a decrease of the
continuum threshold $V_\infty$, as already mentioned above.

\begin{figure}
\includegraphics[width=14cm]{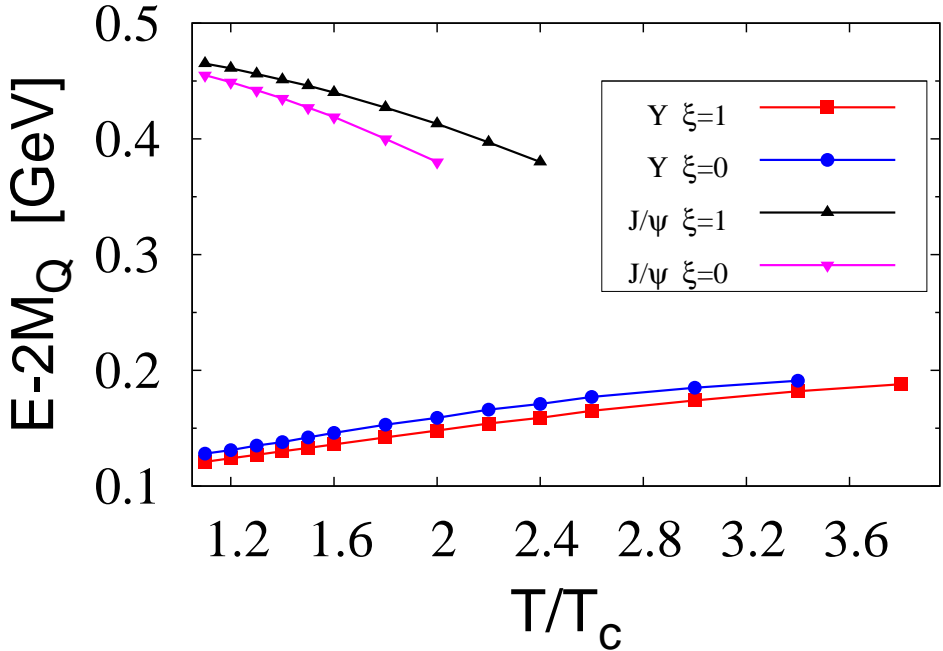}
\vspace*{-1cm}
\caption[a]{Absolute energy of the $J/\Psi$ and $\Upsilon$ states,
  minus twice the corresponding quark mass, as a function of
  temperature.}
\label{fig:E-2m}
\end{figure}
In this vein, it is also instructive to look at the behavior of the
absolute energy of the $J/\Psi$ and $\Upsilon$ states versus
temperature, shown in fig.~\ref{fig:E-2m}. We recall that $E-2m_Q =
\langle V_\infty\rangle + E_{\rm bind}$. The energy of the $\Upsilon$
increases slightly with temperature as is expected for a small-size
state bound mainly by the Debye-Coulomb part of the potential (plus a
constant): the first term on the right hand side of
eq.~(\ref{KMSpot_xi2}) {\em increases} with the screening mass
$\mu$. On the other hand, $E_{J/\Psi}$ decreases with $T$ because the
second term in~(\ref{KMSpot_xi2}) {\em decreases} as $\mu$ increases.

\begin{figure}
\includegraphics[width=14cm]{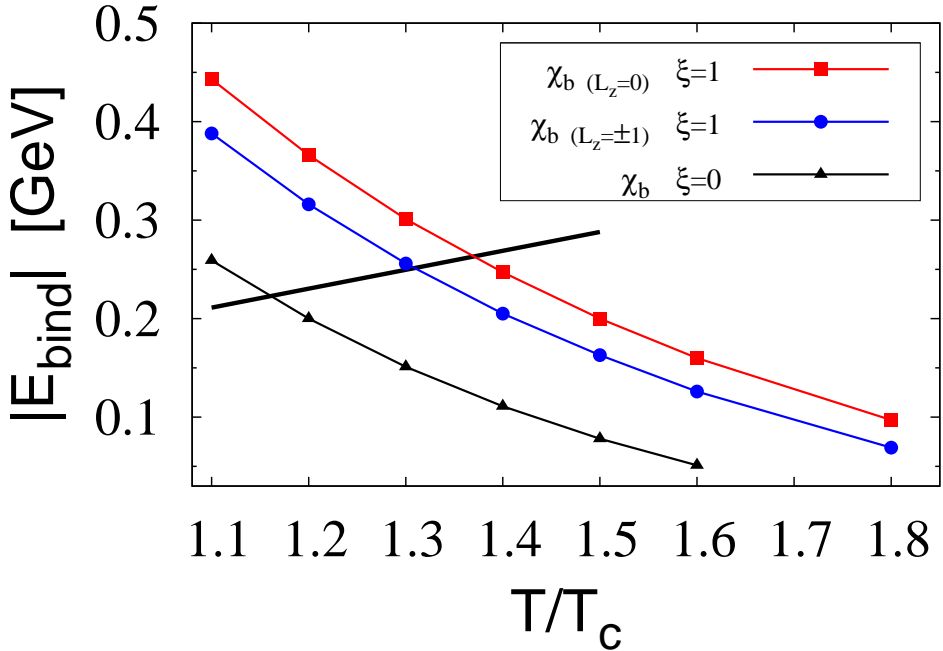}
\vspace*{-1cm}
\caption[a]{Temperature-dependence of the binding energy for the
 $1P$ state of bottomonium for two values of the plasma anisotropy
 parameter $\xi$. The straight line corresponds to a binding energy
 equal to the temperature.}
\label{fig:Eb_bott1P}
\end{figure}
Fig.~\ref{fig:Eb_bott1P} shows the temperature dependence of the
binding energies of the $1P$ states of bottomonium\footnote{The $1P$
  states of charmonium do not have binding energies which exceed the
  temperature significantly for $T\simge T_C$.}, identified with the
$\chi_b$. The anisotropy again leads to an increase of
$|E_{\text{bind}}|$ by about 50\%, comparable to the behavior of the
$J/\Psi$ from above. It also leads to a preferred polarization of the
$\chi_b$, with about 50~MeV splitting between states with angular
momentum $L_z=0$ and $L_z=\pm1$, respectively. At $T\sim T_c$, due to
the Boltzmann factor~(\ref{eq:Boltz}), the population of the state
with $L_z=0$ is about 30\% higher than that of either one of the
$L_z=\pm1$ states. Here, the polarization is with respect to the axis
${\bf n}$ of anisotropy, which coincides with the direction of
expansion of the plasma. In addition, quarkonium states produced in
high-energy collisions initially through semi-hard gluon fusion may
exhibit polarization with respect to the particle velocity
vector~\cite{Ioffe:2003rd}.

\section{Summary and Conclusions} 
\label{sec:Summary}

In a viscous plasma, anisotropic expansion of a fluid element leads to
an anisotropy of the quasi-particle momentum distributions. The Hard
Thermal Loop resummed propagator of static chromo-electric fields then
carries angular dependence which leads to anisotropic Debye screening.

In this paper we have proposed the first model for the static
potential between a very heavy quark and anti-quark in a hot
anisotropic QCD plasma. Conceptually, we assume that the time scale
$\sim 1/|E_{\text{bind}}|$ associated with the bound state is short
compared to the time scales over which the temperature and the
anisotropy evolve.

At distances on the order of the Debye length the potential can be
calculated from perturbative QCD (at high temperature). At larger
distances it is however dominated by the non-perturbative string
attraction. Lattice gauge theory simulations have shown that in the
deconfined phase the string is screened at a similar scale $r_{\rm
  med}(T)\sim 1/m_D(T)$ and that at infinite separation the free
energy of a $Q\bar{Q}$ pair approaches a constant $V_\infty(T)$, equal
to twice the free energy of a single heavy quark in the plasma.

The essential features appear to be in qualitative agreement with a
model proposed by Karsch, Mehr and Satz~\cite{KMS} long time
ago. However, to obtain the heavy-quark potential we subtract the
entropy contribution from their {\em ansatz} for the free energy of a
$Q\bar{Q}$ pair. We thereby obtain the internal energy of the
$Q\bar{Q}$ pair which should be viewed as an upper limit for the
physical potential. The latter may be less binding than the KMS
internal energy used here. We note that in the relevant temperature
region, up to about $2T_c$, the potential at short and intermediate
distances follows the zero-temperature Cornell-potential; i.e., the
overshooting problem of the internal energy is eliminated, in
accordance with lattice data on the free energy.

The KMS model correctly reproduces the Cornell potential
at short distances and, moreover, does not introduce any new
parameters besides the string tension. This is important for our
present goal of extending the isotropic potential to anisotropic
plasmas. Knowledge of the anisotropic screening scale obtained from
the gluon propagator is sufficient to generalize the KMS
model to anisotropic media.

We then proceed to solve the Schr\"odinger equation with this
potential to determine the wave functions of bound $c\bar{c}$ and
$b\bar{b}$ states in the plasma. The radial Schr\"odinger equation is
no longer sufficient as the potential carries angular dependence. We
employ a finite difference time domain method (in Euclidean time) on a
three-dimensional lattice to obtain the wave functions and the binding
energies. Some medium effects are neglected in this approach. However,
solving for the full non-relativistic Green's function
(incl.\ threshold effects) in three dimensions is beyond the scope of
this paper.

We find that just above the critical temperature $T_c\simeq192$~MeV
for deconfinement (in QCD with $N_c=3$ colors) in an anisotropic
medium both the $1S$ state of charmonium as well as the $1S$ and $1P$
states of bottomonium have binding energies larger than $T$; the
temperature may serve as a rough estimate for the width of the states.
The binding energies decrease with temperature and cease to exceed the
estimated width $\Gamma\sim T$ at some higher temperatures. We note,
also, that the Boltzmann enhancement factor $\exp(-E_{\rm bind}/T)$
for bound states is negligible anyways when $|E_{\rm bind}| \simle
T$. 

The decrease of $|E_{\rm bind}|$ with $T$ is due to two effects:
First, the continuum threshold $V_\infty(T)$ decreases approximately
like $\sim1/T$. The energy gap between the bound state and the
continuum, which is the binding energy, therefore decreases, too. In
fact, for the model adopted here, we find this to be the dominant
effect on the $1S$ ground state of bottomonium whose wave function is
rather insensitive to the presence of the medium. The state is too
small to be affected strongly by screening. Hence, the $\Upsilon$ peak
in the dilepton invariant mass distribution may not experience a large
shift although one should expect substantial broadening near the
dissociation temperature.

Larger states such as the $1S$ ground state of charmonium and the $1P$
excited state of bottomonium, however, may also experience some
modifications due to screening. The root-mean-square radii of these
states increase rather rapidly with $T$ around the dissociation point
$|E_{\rm bind}| \sim T$.

The two main results of this work are as follows. At fixed $T$, the
screening mass decreases with increasing $\xi$. In the KMS model, the
asymptotic value of the potential is intrinsically related to the
screening mass via $V_\infty(\theta)\sim 1/\mu(\theta;\xi,T)$. Hence,
less screening translates into an increase of the potential at
infinite separation. This implies that the binding energies of bound
states increase, too. The effect is quite substantial even for
moderate anisotropies $\xi\simeq1$ considered here: we find that just
above $T_c$ the binding energy of the bottomonium ground state
increases by about 30\%, that of $1S$ charmonium and of $1P$
bottomonium by 50\%. Thus, such quarkonium states may exist up to
somewhat higher temperatures than in case of an isotropic, perfectly
equilibrated medium (for $\xi=0$ the $J/\psi$ and the $\Upsilon$ are
expected to dissociate by $1.2\,T_c$ and $1.8\,T_c$, respectively, in
agreement with previous potential model calculations).

The other important new effect identified here is that the angular
dependence of the inter-quark potential in an anisotropic medium
induces a polarization of states with non-zero angular
momentum. According to our estimates, the splitting of the $\chi_b$
with $L_z=0$ and $L_z=\pm1$, respectively, is on the order of
50~MeV. At $T\simeq200$~MeV, the population of the state with $L_z=0$
is Boltzmann-enhanced by about 30\% as compared to the states with
angular momentum along the direction of anisotropy, respectively. The
experimental confirmation of such a polarization at RHIC or LHC may
provide first evidence for a non-zero viscosity of QCD near
$T_c$. 

The next step of our investigation is the determination of the
imaginary part of the potential, which will provide insight into how
the anisotropy of the medium affects the widths of the states. We
shall present results in a future publication.

\section*{Acknowledgments}
We thank D.\ Kharzeev and P.\ Petreczky for reading the manuscript
prior to publication and for useful comments. Y.G.\ thanks the
Helmholtz foundation and the Otto Stern School at Frankfurt university
for their support and the center for scientific computing (CSC) for
computational resources.



\begin{thebibliography}{99}

\bibitem{Lucha:1991vn}
W.~Lucha, F.~F.~Schoberl and D.~Gromes,
Phys.\ Rept.\  {\bf 200}, 127 (1991).

\bibitem{Eichten:1979ms}
E.~Eichten, K.~Gottfried, T.~Kinoshita, J.~B.~Kogut, K.~D.~Lane and T.~M.~Yan,
Phys.\ Rev.\ Lett.\  {\bf 34}, 369 (1975)
[Erratum-ibid.\  {\bf 36}, 1276 (1976)];
E.~Eichten, K.~Gottfried, T.~Kinoshita, K.~D.~Lane and T.~M.~Yan,
Phys.\ Rev.\  D {\bf 17}, 3090 (1978)
[Erratum-ibid.\  D {\bf 21}, 313 (1980)];
Phys.\ Rev.\  D {\bf 21}, 203 (1980).

\bibitem{Brambilla:2004jw}
  A.~Pineda and J.~Soto,
  Nucl.\ Phys.\ Proc.\ Suppl.\  {\bf 64}, 428 (1998);
N.~Brambilla, A.~Pineda, J.~Soto and A.~Vairo,
Rev.\ Mod.\ Phys.\  {\bf 77}, 1423 (2005)
[arXiv:hep-ph/0410047].

\bibitem{GPY}
E.~V.~Shuryak,
Phys.\ Rept.\  {\bf 61}, 71 (1980);
D.~J.~Gross, R.~D.~Pisarski and L.~G.~Yaffe,
Rev.\ Mod.\ Phys.\  {\bf 53}, 43 (1981).

\bibitem{MatsuiSatz}
T.~Matsui and H.~Satz,
Phys.\ Lett.\  B {\bf 178}, 416 (1986).

\bibitem{KMS}
F.~Karsch, M.~T.~Mehr and H.~Satz,
Z.\ Phys.\  C {\bf 37}, 617 (1988).

\bibitem{mocsy}
  \'A.~M\'ocsy and P.~Petreczky,
  Phys.\ Rev.\  D {\bf 73}, 074007 (2006)
  [arXiv:hep-ph/0512156];
  Eur.\ Phys.\ J.\  C {\bf 43}, 77 (2005)
  [arXiv:hep-ph/0411262].

\bibitem{wong}
  C.~Y.~Wong,
  Phys.\ Rev.\  C {\bf 72}, 034906 (2005)
  [arXiv:hep-ph/0408020].
  
\bibitem{alberico}
  W.~M.~Alberico, A.~Beraudo, A.~De Pace and A.~Molinari,
  Phys.\ Rev.\  D {\bf 75}, 074009 (2007)
  [arXiv:hep-ph/0612062];
  Phys.\ Rev.\  D {\bf 77}, 017502 (2008)
  [arXiv:0706.2846 [hep-ph]].

\bibitem{rapp}
  D.~Cabrera and R.~Rapp,
  Phys.\ Rev.\  D {\bf 76}, 114506 (2007)
  [arXiv:hep-ph/0611134].

\bibitem{Mocsy:2007jz}
 \'A.~M\'ocsy  and P.~Petreczky,
Phys.\ Rev.\ Lett.\  {\bf 99}, 211602 (2007)
[arXiv:0706.2183 [hep-ph]].

\bibitem{Mocsy:2007yj}
 \'A.~M\'ocsy  and P.~Petreczky,
Phys.\ Rev.\  D {\bf 77}, 014501 (2008)
[arXiv:0705.2559 [hep-ph]].

\bibitem{datta}
S.~Datta, F.~Karsch, P.~Petreczky and I.~Wetzorke,
Phys.\ Rev.\ D {\bf 69}, 094507 (2004)
[arXiv:hep-lat/0312037].

\bibitem{jakovac}
  A.~Jakov\'ac, P.~Petreczky, K.~Petrov and A.~Velytsky,
  Phys.\ Rev.\  D {\bf 75}, 014506 (2007)  
[arXiv:hep-lat/0611017].

\bibitem{aarts}
 G.~Aarts {\it et al .},
  Phys.\ Rev.\  D {\bf 76}, 094513 (2007)
  [arXiv:0705.2198 [hep-lat]].

\bibitem{umeda}
  T.~Umeda,
  Phys.\ Rev.\  D {\bf 75}, 094502 (2007)
  [arXiv:hep-lat/0701005].

\bibitem{Mocsy:2008eg}
  \'A.~M\'ocsy,
  arXiv:0811.0337 [hep-ph], to appear in EPJC.

\bibitem{LHCpred}
N.~Armesto {\it et al.},
J.\ Phys.\ G {\bf 35}, 054001 (2008)
[arXiv:0711.0974 [hep-ph]];
R.~Rapp, D.~Blaschke and P.~Crochet,
arXiv:0807.2470.

\bibitem{laine}
  M.~Laine, O.~Philipsen and M.~Tassler,
  JHEP {\bf 0709}, 066 (2007)
  [arXiv:0707.2458 [hep-lat]];
  M.~Laine,
  JHEP {\bf 0705}, 028 (2007)
  [arXiv:0704.1720 [hep-ph]].

\bibitem{blaizot}
  A.~Beraudo, J.~P.~Blaizot and C.~Ratti,
  Nucl.\ Phys.\  A {\bf 806}, 312 (2008)
  [arXiv:0712.4394 [nucl-th]].

\bibitem{Brambilla:2008cx}
N.~Brambilla, J.~Ghiglieri, A.~Vairo and P.~Petreczky,
Phys.\ Rev.\  D {\bf 78}, 014017 (2008)
[arXiv:0804.0993 [hep-ph]];
M.~A.~Escobedo and J.~Soto,
arXiv:0804.0691 [hep-ph].

\bibitem{Dumitru:2007hy}
A.~Dumitru, Y.~Guo and M.~Strickland,
Phys.\ Lett.\  B {\bf 662}, 37 (2008)
[arXiv:0711.4722 [hep-ph]];
  Y.~Guo,
  arXiv:0809.3873 [hep-ph].

\bibitem{Romatschke:2003ms}
P.~Romatschke and M.~Strickland,
Phys.\ Rev.\ D {\bf 68}, 036004 (2003) [arXiv:hep-ph/0304092].

\bibitem{Mrowczynski:2004kv}
  S.~Mrowczynski, A.~Rebhan and M.~Strickland,
  Phys.\ Rev.\  D {\bf 70}, 025004 (2004)
  [arXiv:hep-ph/0403256].

\bibitem{Romatschke:2004jh}
  P.~Romatschke and M.~Strickland,
  Phys.\ Rev.\  D {\bf 70}, 116006 (2004)
  [arXiv:hep-ph/0406188].

\bibitem{Schenke:2006fz}
  B.~Schenke and M.~Strickland,
  Phys.\ Rev.\  D {\bf 74}, 065004 (2006)
  [arXiv:hep-ph/0606160].

\bibitem{Asakawa:2006tc}
eq.\ (6-40) in M.~Asakawa, S.~A.~Bass and B.~M\"uller,
Prog.\ Theor.\ Phys.\  {\bf 116}, 725 (2007)
[arXiv:hep-ph/0608270].

\bibitem{Petreczky:2005bd}
  P.~Petreczky,
  Eur.\ Phys.\ J.\  C {\bf 43}, 51 (2005)
  [arXiv:hep-lat/0502008].
  
\bibitem{Shuryak:2004tx}
E.~V.~Shuryak and I.~Zahed,
Phys.\ Rev.\  D {\bf 70}, 054507 (2004)
[arXiv:hep-ph/0403127].

\bibitem{kaczmarekHP} 
  O.~Kaczmarek,
  PoS C {\bf POD07}, 043 (2007)
  [arXiv:0710.0498 [hep-lat]].

\bibitem{Kaczmarek:2004gv}
O.~Kaczmarek, F.~Karsch, F.~Zantow and P.~Petreczky,
Phys.\ Rev.\ D {\bf 70}, 074505 (2004)
[Erratum-ibid.\ D {\bf 72}, 059903 (2005)]
[arXiv:hep-lat/0406036].

\bibitem{Megias:2007pq}
E.~Megias, E.~Ruiz Arriola and L.~L.~Salcedo,
JHEP {\bf 0601}, 073 (2006)
[arXiv:hep-ph/0505215];
Phys.\ Rev.\  D {\bf 75}, 105019 (2007).
[arXiv:hep-ph/0702055].

\bibitem{Megias:2009mp}
E.~Megias, E.~R.~Arriola and L.~L.~Salcedo,
arXiv:0903.1060 [hep-ph].

\bibitem{Bali}
G.~S.~Bali, K.~Schilling and A.~Wachter,
Phys.\ Rev.\  D {\bf 56}, 2566 (1997)
[arXiv:hep-lat/9703019].

\bibitem{MTB}
R.~Baier and Y.~Mehtar-Tani,
Phys.\ Rev.\  C {\bf 78}, 064906 (2008).

\bibitem{Sudiarta:2007}
I.~W.~Sudiarta and D.~J.~W. Geldart,
Journal of Physics A {\bf 40} 8, 1885 (2007).

\bibitem{Kaczmarek:2005}
O.~Kaczmarek and F.~Zantow,
Phys.\ Rev.\ D {\bf 71}, 114510 (2005) [arXiv:hep-lat/0503017].

\bibitem{dima}
  D.~Kharzeev, L.~D.~McLerran and H.~Satz,
  Phys.\ Lett.\  B {\bf 356}, 349 (1995)
  [arXiv:hep-ph/9504338].

\bibitem{Ioffe:2003rd}
B.~L.~Ioffe and D.~E.~Kharzeev,
Phys.\ Rev.\  C {\bf 68}, 061902 (2003)
[arXiv:hep-ph/0306176].



\end{thebibliography}
\end{document}